\documentclass[aps,preprint,nofootinbib,floatfix]{revtex4-2}
\usepackage[utf8]{inputenc}

\usepackage[title]{appendix}
\usepackage[colorlinks,
linkcolor=blue,
filecolor=blue,
anchorcolor=blue,
urlcolor=blue,
citecolor=blue,
bookmarks=false,
]{hyperref}
\usepackage{graphicx}
\usepackage{amssymb}
\usepackage{amsmath}
\usepackage{subfigure}
\usepackage{bm}
\usepackage{booktabs}
\usepackage{xcolor}
\usepackage{color}

\allowdisplaybreaks[4]
\begin{document}
\title{\boldmath Higgs Decays to $Z\gamma$ and $\gamma\gamma$ in the Flavor-Gauged Two Higgs Doublet Model}
\author{Feng-Zhi Chen}
\author{Qiaoyi Wen}
\author{Fanrong Xu}
\email{fanrongxu@jnu.edu.cn}

\affiliation{\it Department of Physics, College of Physics $\&$ Optoelectronic Engineering,\\
Jinan University, Guangzhou 510632, P.R. China}


\begin{abstract}

This work examines the $h\to Z\gamma$ and $h\to\gamma\gamma$ decays in the flavor-gauged two Higgs doublet model (FG2HDM), which augments the Standard Model (SM) with an additional scalar doublet, a singlet, and a $U(1)'$ flavor gauge symmetry. Beyond the SM spectrum, FG2HDM predicts five additional physical scalars and a new neutral gauge boson, $Z'$. We demonstrate that while both decay channels are sensitive to charged Higgs loops, $h \to Z\gamma$ is uniquely modified by fermion-antifermion-$Z$ ($f\bar{f}Z$) vertex corrections. These vertex corrections further impact top-quark observables and the flavor-changing neutral current (FCNC) process $b\to s\ell^+\ell^-$. Our analysis identifies a viable parameter space ($m_{H^\pm}>200$~GeV and $\lambda_{hH^+H^-}<0$) consistent with current $1\sigma$ experimental limits, where the signal strength $\mu_{\gamma\gamma}$ remains the primary constraint on scalar sector parameters. Regarding the $f\bar{f}Z$ couplings, we delineate the allowed regions in the $\mathcal{Q}_{tL}$-$\mathcal{Q}_{tR}$ plane by evaluating the leading top-quark contributions, revealing that $b\to s\ell^+\ell^-$ imposes the most stringent bounds. Finally, we highlight that the $14\%$ projected precision for $\mu_{Z\gamma}$ at the High-Luminosity LHC (HL-LHC) will significantly enhance sensitivity to the FG2HDM.
 
\end{abstract}

\maketitle
\newpage

\section{Introduction}
\label{sec:intro}

The discovery of the Higgs boson at the Large Hadron Collider (LHC) by the ATLAS and CMS collaborations in 2012 marked a significant triumph for the SM~\cite{ATLAS:2012yve,CMS:2012qbp,CMS:2013btf}. Despite the significant progress made in particle physics, several fundamental questions remain unanswered, such as the matter-antimatter asymmetry in the universe, the origin of neutrino mass, and the nature of dark matter. These unsolved mysteries underscore the limitations of the SM. Since then, the quest for new physics (NP) beyond the SM (BSM) has been one of the central endeavors in elementary particle physics. 

One attractive extension to the SM, which involves adding an additional Higgs doublet, was first proposed by T. D. Lee to introduce the spontaneous CP violation~\cite{Lee:1973iz}, and is now dubbed as the two Higgs doublet model (2HDM).  However, a major drawback of the generic 2HDM is that it typically introduces the tree-level fFCNC in Yukawa interactions that are unexpected in a model construction. FCNC generation is suppressed by raising various modified 2HDMs to control the two Yukawa interactions. This includes imposing different $Z_2$ charges for the two scalars~\cite{Glashow:1976nt,Paschos:1976ay}, aligning the two Yukawa matrices~\cite{Pich:2009sp,Jung:2010ik,Ferreira:2010xe}, or making the two Yukawa interactions proportional to the off-diagonal elements of the Cabibbo–Kobayashi–Maskawa (CKM) matrix~\cite{Cabibbo:1963yz,Kobayashi:1973fv}. The last one is usually called the Branco-Grimus-Lavoura (BGL) model, named after the authors of this model~\cite{Branco:1996bq,Branco:1996bq}. A flavor gauged version of the BGL model, i.e., FG2HDM, was first proposed in Ref.~\cite{Celis:2015ara} and was further developed in Refs.~\cite{Ordell:2019zws,Ordell:2020yoq,Ferreira:2022zil}. FG2HDM extends the original 2HDM by incorporating a scalar singlet field and an additional $U(1)'$ flavor gauge symmetry. Compared with the SM particle spectrum, the FG2HDM features five additional physical scalars and one neutral gauge boson $Z'$. 

Recently, the ATLAS and CMS collaborations reported initial evidence for the $h\to Z\gamma$ decay.  The observed signal strength\footnote{The signal strength $\mu_{X}$ (where $X$ is any Higgs decay product) is defined as the product of the cross section and the branch fraction ($\sigma(pp\to h) \mathcal{B}(h\to X)$) relative to the SM prediction. The cross section aligns well with the SM prediction, so we assumed that the NP merely affect the branching fractions.} $\mu_{Z\gamma}$ was measured at the $68\%$ confidence level (CL) to be $2.0_{-0.9}^{+1.0}$ by ATLAS~\cite{ATLAS:2020qcv}, $2.4\pm{0.9}$ by CMS~\cite{CMS:2022ahq}, and $2.2\pm0.7$ in their combined analysis~\cite{ATLAS:2023yqk}. Such a signal strength deviates from the SM prediction $\mu_{Z\gamma}^\mathrm{SM}=1$ by $1.9\sigma$, motivating many possible explanations from two distinct perspectives. First, as $h\to Z\gamma$ is a loop-induced process, higher-order corrections to the leading-order (LO) amplitudes from one-loop amplitudes are crucial.\footnote{Refer to Ref.~\citep{Khodjamirian:2025fox} for an estimated magnitude of leading non-perturbative QCD corrections ($10^{-5}$).} These include two-loop QCD corrections~\cite{Spira:1991tj,Gehrmann:2015dua,Bonciani:2015eua,Buccioni:2023qnt}, providing a $0.3\%$ increase, and two-loop electroweak corrections~\cite{Chen:2024vyn,Sang:2024vqk}, yielding a $7\%$ enhancement. Clearly, these next-to-leading-order (NLO) corrections alone cannot account for the $1.9\sigma$ excess. Alternatively, the deviation may stem from BSM contributions, a possibility extensively explored in the literature (e.g.,~\cite{Panghal:2023iqd,Lichtenstein:2023vza,Hong:2023mwr,Boto:2023bpg,Das:2024tfe,Cheung:2024kml,He:2024bxi,Ouazghour:2024fgo,Arhrib:2024wjj,Hernandez-Juarez:2024iwe,Grojean:2024tcw,Hu:2024slu,Arhrib:2024itt,Phan:2024jbx,Israr:2024ubp,Mantzaropoulos:2024vpe,Boto:2024tzp,Barducci:2023zml}). However, integrating the most recent ATLAS measurement ($\mu_{Z\gamma}=0.9^{+0.7}_{-0.6}$)~\cite{ATLAS:2025qvd} with previous results yields a revised signal strength of $\mu_{Z\gamma}=1.3^{+0.6}_{-0.5}$, bringing the observation into agreement with SM expectations. While this reduces the immediate tension, analysing $h\to Z\gamma$ within specific BSM frameworks still requires a simultaneous consideration of the $h\to\gamma\gamma$ channel. This is due to the inherent correlation between the two, as both are mediated by the same charged loop-particles~\cite{Gunion:1989we,Djouadi:2005gi}. Given that the current experimental value $\mu_{\gamma\gamma}=1.10\pm0.06$~\cite{ParticleDataGroup:2024cfk}, highly accords with the SM, it imposes stringent constraints on the FG2HDM parameter space, effectively narrowing the regions permitted by the $\mu_{Z\gamma}$ data.

Building on our previous work~\cite{Wen:2025smu}, where specific Yukawa matrix textures in FG2HDM were proposed to explain the long-standing $B$ anomalies, the present study investigates whether the model's parameter space remains viable under the current experimental constraints on $\mu_{Z\gamma}$ and $\mu_{\gamma\gamma}$. While the $h\to Z\gamma$ and $h\to\gamma\gamma$ channels have been explored in other frameworks such as the complex 2HDM (C2HDM)~\cite{Fontes:2014xva} using earlier data~\cite{ATLAS:2014fxe,CMS:2013rmy} the recent updates from ATLAS and CMS~\cite{ATLAS:2020qcv,CMS:2022ahq,ATLAS:2023yqk,ATLAS:2025qvd} necessitate a more refined numerical and theoretical analysis. In FG2HDM, both decay modes receive loop-level contributions from charged Higgs bosons; furthermore, $h \to Z\gamma$ is modified by corrections to the $f\bar{f}Z$ vertex. Consequently, this work aims to identify the permitted FG2HDM parameter space for $\mu_{Z\gamma}$ and $\mu_{\gamma\gamma}$, while accounting for the constraints on the $f\bar{f}Z$ vertex from top-quark measurements and the $b\to s\ell^+\ell^-$ FCNC process.

The remainder of the paper is organized as follows: In Sec.~\ref{sec:model}, we provide a brief overview of the key features of the FG2HDM. In Sec.~\ref{sec:oneloop}, we calculate the one-loop amplitudes for the $h\to Z\gamma$ and $h\to\gamma\gamma$ decays in the unitary gauge. The numerical analysis and subsequent discussions are detailed in Sec.~\ref{sec:num}. Our conclusions are presented in Sec.~\ref{sec:con}. For convenience, the relevant Lagrangian and the corresponding Feynman rules are prepared in Appendix~\ref{app:La}. In addition, the explicit expressions for the scalar functions obtained from one-loop calculations are listed in Appendix~\ref{app:scalarfunc}, along with the kinematics for Higgs decays in Appendix~\ref{app:kin}.

\section{The model}
\label{sec:model}

FG2HDM is a specific 2HDM comprising two $SU(2)_L$ scalar doublets, $\Phi_1$ and $\Phi_2$, and a scalar singlet $S$, and endowed with an additional $U(1)'$ flavor gauge symmetry. Here, the $U(1)'$ charges, denoted as $X_i$ ($i=L_L,\cdots,S$), are assigned to ensure the $U(1)'$ gauge invariance. The detailed relationships among these $U(1)'$ charges for various fields are elaborated in Appendix~A of Ref.~\cite{Wen:2025smu}. In the presence of the new $U(1)'$ gauge symmetry, the gauge anomaly constraints should be considered further. Note that various anomaly-free FG2HDMs have been surveyed and studied in Refs.~\cite{Ordell:2019zws,Ordell:2020yoq,Ferreira:2022zil}. 

For convenience, we collect the quantum representations for various fields in FG2HDM under the $SU(2)_L\times U(1)_Y\times U(1)'$ symmetries in Table~\ref{tab:quantum}. With these assignments, the total Lagrangian of FG2HDM can be expressed as~\cite{Wen:2025smu}
\begin{align}
\mathcal{L}=\mathcal{L}_S+\mathcal{L}_G+\mathcal{L}_F\,,
\end{align}
where subscripts $S$, $G$, and $F$ denote respectively for the scalar, gauge, and fermion components of the total Lagrangian:\footnote{In principle, Eq.~\eqref{eq:gauge} could include a gauge-symmetry-allowed kinetic mixing term, $-\frac{\sin\epsilon}{2}B_{\mu\nu}\hat{Z}^{\prime\mu\nu}$, which would primarily result in a redefinition of the $U(1)'$ charges. However, since its effect is to shift the constraints from the original $U(1)'$ charges to these redefined ones, without altering the numerical results or physical predictions of the model, we can safely neglect this term for simplicity.}
\begin{align}
\mathcal{L}_S&=D^\mu\Phi_1^\dagger D_\mu\Phi_1+D^\mu\Phi_2^\dagger D_\mu\Phi_2+D^\mu S^\ast D_\mu S-V(\Phi_1,\Phi_2,S)\,,\label{eq:scalar}\\[0.2cm]
{\color{blue} \mathcal{L}_G}&{\color{blue}=-\frac{1}{4}W^a_{\mu\nu} W^{a\mu\nu}-\frac{1}{4}B_{\mu\nu} B^{\mu\nu}-\frac{1}{4}\hat{Z}^\prime_{\mu\nu}\hat{Z}^{\prime\mu\nu}\,,}\label{eq:gauge}\\[0.2cm]
\mathcal{L}_F&=\bar{Q}_L iD_\mu\gamma^\mu Q_L+\bar{L}_L iD_\mu\gamma^\mu L_L+\bar{u}_R iD_\mu\gamma^\mu u_R+\bar{d}_R iD_\mu\gamma^\mu d_R+\bar{e}_R iD_\mu\gamma^\mu e_R\notag\\
&-\left[\bar{Q}_L(Y_1^d\Phi_1+Y_2^d\Phi_2)d_R+\bar{Q}_L(Y_1^u\tilde{\Phi}_1+Y_2^u\tilde{\Phi}_2)u_R+\bar{L}_L(Y_1^\ell\Phi_1+Y_2^\ell\Phi_2)e_R+\mathrm{h.c.}\right]\,.\label{eq:fermion}
\end{align}
where $L_L=(\nu_L,e_L)^T$ and $Q_L=(u_L,d_L)^T$ represent the left-handed lepton and quark $SU(2)_L$ doublets, respectively, and $e_R$, $u_R$, and $d_R$ denote the right-handed lepton, up-type quark, and down-type quark $SU(2)_L$ singlets, respectively. The physical significance of each component is elaborated subsequently.
\begin{table}[tbh!]
\centering
\tabcolsep 0.15in
\begin{tabular}{|c|cccccccc|}
\hline
 Fields & $L_L$ & $e_R$ & $Q_L$ & $u_R$ & $d_R$ & $\Phi_1$ & $\Phi_2$ & $S$  \\
\hline
$SU(3)_C$ & \textbf{1} & \textbf{1} & \textbf{3} & \textbf{3} & \textbf{3} & \textbf{1} & \textbf{1} & \textbf{1} \\
$SU(2)_L$ & \textbf{2} & \textbf{1} & \textbf{2} & \textbf{1} & \textbf{1} & \textbf{2} & \textbf{2} & \textbf{1} \\
$U(1)_Y$ & $-\frac{1}{2}$ & $-1$ & $\frac{1}{6}$ & $\frac{2}{3}$ & $-\frac{1}{3}$ & $\frac{1}{2}$ & $\frac{1}{2}$ & $0$ \\
$U(1)'$ & $X_{L_L}$ & $X_{e_R}$ & $X_{Q_L}$ & $X_{u_R}$ & $X_{d_R}$ & $X_{\Phi_1}$ & $X_{\Phi_2}$ & $X_{S}$ \\
\hline
\end{tabular}
\caption{\small The quantum numbers of different fields under the $SU(3)_C\times SU(2)_L\times U(1)_Y\times U(1)'$ symmetries.}
\label{tab:quantum}
\end{table}

The scalar potential in Eq.~\eqref{eq:scalar} is given by~\cite{Wen:2025smu}
\begin{align}\label{eq:potential}
 V(\Phi_1,\Phi_2,S) &= m_{11}^2 \Phi_1^\dagger\Phi_1 + m_{22}^2 \Phi_2^\dagger\Phi_2
    + \frac{\lambda_1}{2} (\Phi_1^\dagger\Phi_1)^2 + \frac{\lambda_2}{2} (\Phi_2^\dagger\Phi_2)^2 \nonumber\\
    & + \lambda_3 (\Phi_1^\dagger\Phi_1)(\Phi_2^\dagger\Phi_2) + \lambda_4 (\Phi_1^\dagger\Phi_2)(\Phi_2^\dagger\Phi_1) \nonumber\\
    & + m_S^2 |S|^2 + \lambda_S |S|^4 + \kappa_1 |\Phi_1|^2 |S|^2 + \kappa_2 |\Phi_2|^2 |S|^2 \nonumber\\
    & + \kappa_3 \left(\Phi_1^\dagger\Phi_2 S^2 + \Phi_2^\dagger\Phi_1 (S^\ast)^2\right)\,,
\end{align}
where $m_{11,22,S}$, $\lambda_{1,2,3,4,S}$, and $\kappa_{1,2,3}$ denote the coupling constants of various interaction terms. The two doublets and the singlet in the flavor basis are parametrized as
\begin{align}\label{eq:parametrize}
\Phi_1=\begin{pmatrix}
\phi_1^+\\
\frac{1}{\sqrt{2}}(\rho_1+i\eta_1+v_1)
\end{pmatrix}\,,\quad 
\Phi_2=\begin{pmatrix}
\phi_2^+\\
\frac{1}{\sqrt{2}}(\rho_2+i\eta_2+v_2)
\end{pmatrix}\,,\quad 
S=\frac{1}{\sqrt{2}}(s_0+i\chi_0+v_S)\,,
\end{align}
where the vacuum expectation values (VEVs) $v_1$ and $v_2$ combine to form the SM Higgs VEV $v=1/(\sqrt{2}G_F)^{1/2}=\sqrt{v_1^2+v_2^2}\simeq246$~GeV, with $G_F$ being the Fermi constant. The singlet VEV $v_S$ is introduced to spontaneously break the $U(1)'$ symmetry. Rotating to the mass basis yields
\begin{align}\label{eq:rota}
        &\begin{pmatrix}
            G^\pm\\
            H^\pm
        \end{pmatrix}=U_{1}\begin{pmatrix}
        \phi_1^\pm\\
        \phi_2^\pm
    \end{pmatrix}\,,\quad
        \begin{pmatrix}
            G^{0}\\
            G^{\prime0}\\
            H_A
        \end{pmatrix}=U_{2}\begin{pmatrix}
            \eta_1\\
            \eta_2\\
            \chi_0
        \end{pmatrix}\,,\quad
        \begin{pmatrix}
            H\\
            h\\
            H_S
        \end{pmatrix}=U_{3}\begin{pmatrix}
            \rho_1\\
            \rho_2\\
            s_0
        \end{pmatrix}\,,
\end{align}
where $U_{1}$, $U_{2}$, and $U_{3}$ are the rotation matrices defined as
\begin{align}
		U_{1}\equiv
		\begin{pmatrix}
			\cos\beta&\sin\beta\\
			-\sin\beta&\cos\beta
		\end{pmatrix}\,,\quad
		U_{2}\equiv\begin{pmatrix}
			1&0\\
			0&U_\gamma
		\end{pmatrix}\begin{pmatrix}
			U_1&0\\
			0&1
		\end{pmatrix}\,,\quad
		U_{3}\equiv\begin{pmatrix}
			-\cos\alpha&-\sin\alpha&0\\
			\sin\alpha&-\cos\alpha&0\\
			0&0&1
		\end{pmatrix}\,.
	\end{align}
Here, $\alpha$ and $\beta$ are the rotation angles, and $U_\gamma$ and $\tan2\alpha$ are given by
\begin{align}
U_\gamma=\begin{pmatrix}
		\frac{2v_1v_2}{\sqrt{4v_1^2v_2^2+v^2v_S^2}}&\frac{-vv_S}{\sqrt{4v_1^2v_2^2+v^2v_S^2}}
		\\
		\frac{vv_S}{\sqrt{4v_1^2v_2^2+v^2v_S^2}}&\frac{2v_1v_2}{\sqrt{4v_1^2v_2^2+v^2v_S^2}}
	\end{pmatrix}\,,\qquad
	\tan2\alpha=\frac{2l}{m-n}\,,
\end{align}
where
\begin{align}
    m = \lambda_1v_1^2 - \frac12\kappa_3\frac{v_2}{v_1}v_S^2\,, \quad
    n = \lambda_2v_2^2 - \frac12\kappa_3\frac{v_1}{v_2}v_S^2\,, \quad
    l = \lambda_{34}v_1v_2 + \frac12\kappa_3v_S^2\,.
\end{align}
After spontaneous symmetry breaking, the four undesired Goldstone bosons $G^\pm$, $G^0$, and $G^{\prime0}$ in Eq.~\eqref{eq:rota} are ``eaten'' to give masses to the $W^\pm$, $Z$, and $Z'$ gauge bosons, respectively. This leaves us with six physical scalars: two charged Higgs $H^\pm$; three neutral scalars $H$, $h$, and $H_S$; and one neutral pseudoscalar $H_A$. Their masses are separately given by
\begin{align}
m_{h,H}^2 = &\frac12 \left[\left(m+n\right) \mp \sqrt{\left(m-n\right)^2 + 4l^2}\right]\,, & m_{H_S}^2 =& 2\lambda_Sv_S^2\,,\notag\\[0.2cm]
m_{H_A}^2 =& -\frac12\kappa_3\left(4v_1v_2+\frac{v^2v_S^2}{v_1v_2}\right)\,, &
m^2_{H^\pm} =& -\frac{1}{2}(\lambda_4 v^2 + \frac{\kappa_3v^2v_S^2}{v_1v_2})\,.
\end{align}
Among the four neutral particles, $h$, $H$, and $H_S$ are CP-even, while $H_A$ is CP-odd. Notably, the scalar $h$ is identified as the Higgs boson discovered at the LHC in 2012~\cite{ATLAS:2012yve,CMS:2012qbp,CMS:2013btf}. 

In Eqs.~\eqref{eq:scalar} and Eq.~\eqref{eq:fermion}, the covariant derivatives for the $SU(2)_L$ doublets and singlets are defined, respectively, as follows:
\begin{align}
        D_\mu &= \partial_\mu - i g_1 Y_i B_\mu - i g' X_{i} \hat{Z}'_\mu - ig_2 \frac{\tau^a}{2} W^a_\mu \,,\label{eq:covd}\\[0.2cm]
        D_\mu &= \partial_\mu - i g_1 Y_i B_\mu - i g' X_{i} \hat{Z}'_\mu\,,\label{eq:covs}
\end{align}
where $\tau^a$ ($a=1,2,3$) is the Pauli matrix, and $B_\mu$, $\hat{Z}'_\mu$, and $W^a_\mu$ are the $U(1)_Y$, $U(1)'$ and $SU(2)_L$ gauge bosons, respectively. Here, $g_1$, $g'$, and $g_2$ are the corresponding coupling constants, and $Y_i$ and $X_{i}$ denote the hypercharge and $U(1)'$ charge of the fields, respectively. After spontaneous symmetry breaking, the gauge bosons acquire their masses. The transformation from the flavor to mass basis is achieved through the mixing matrix, as follows:
\begin{align}
\left(\begin{array}{c}
		A \\ Z \\ {Z}' \end{array}\right)
	= U  \left(\begin{array} {c}
		B\\ W^3 \\ \hat{Z}' \end{array}\right)\,, \quad
	U=\left(\begin{array}{ccc}
		\cos\theta_{W} & \sin\theta_{W} & 0 \\
		-\sin\theta_{W}\cos\theta_{2}^\prime & \cos\theta_{W} \cos\theta_{2}^\prime& \sin\theta_{2}^\prime\\
		\sin\theta_{W}\sin\theta_{2}^\prime& -\cos\theta_{W}\sin\theta_{2}^\prime & \cos\theta_{2}^\prime
	\end{array}\right)\,,
\end{align}
where $\theta_W$ is the Weinberg angle, and $\theta_2^\prime$ the mixing angle between $Z$ and $Z'$. In the limit where $\theta_2^\prime$ approaches zero, the mixing restores to the case in the SM. This arrangement introduces the tree-level FCNCs in the down-type quark sector, which can account for the anomalies observed in $B$ physics~\cite{Wen:2025smu}. In addition, it generates off-diagonal interactions, with the Higgs and $Z$ boson coupling to the charged Higgs and $W$ boson, as well as new corrections to the $f\bar{f}Z$ couplings, thereby causing FG2HDM to contribute more to $h\to Z\gamma$ than $h\to\gamma\gamma$. 

The interactions between fermions and gauge bosons are derived from the covariant derivative terms in the first line of Eq.~\eqref{eq:fermion}. These terms encapsulate how fermions couple to the gauge bosons associated with the symmetries of the model. Meanwhile, the remaining terms in Eq.~\eqref{eq:fermion} depict the Yukawa interactions between fermions and the scalar doublets.\footnote{Note that in these Yukawa interactions, the right-handed charged leptons and down-type quarks couple directly to $\Phi_i$. In contrast, the right-handed up-type quarks couple to the charge-conjugate fields, $\tilde{\Phi}_i\equiv i\tau^2\Phi_i^\ast$.} After spontaneous symmetry breaking, the scalar fields acquire VEVs, which give masses to the fermions.

Armed with the Lagrangian of FG2HDM, we can derive the relevant Feynman rules. In this context, we concentrate solely on the crucial terms indispensable for computing the one-loop amplitudes of the $h\to Z\gamma$ and $h\to\gamma\gamma$ decays. These vital terms, together with their corresponding Feynman rules are cataloged in Appendix~\ref{app:La}.

\section{One-loop amplitudes}
\label{sec:oneloop}

Following a general Lorentz decomposition while enforcing photon gauge invariance, as detailed in Ref.~\cite{Bonciani:2015eua}, we can express the amplitudes for the processes $h(p)\to Z(p_1)\gamma(p_2)$ and $h(p)\to\gamma(p_1)\gamma(p_2)$ in the following form:
\begin{align}\label{eq:Lorentz}
\mathcal{M}_{X}=&(p_2^\mu p_1^\nu-p_1\cdot p_2g^{\mu\nu})\,T_{X}\,\epsilon_{\mu}^\ast(p_1)\epsilon_{\nu}^\ast(p_2)\,,\qquad X=Z\gamma,\gamma\gamma\,,
\end{align}
where $\epsilon_{\mu(\nu)}$ represents the polarization vector of the $Z$ boson or photon, and $T_{X}$ denotes the one-loop functions to be determined. In the presence of NP, the function $T_{X}$ can be written as $T_{X}=T_{X}^\mathrm{SM}+T_{X}^\mathrm{NP}$, and the signal strength $\mu_{X}$ is defined as
\begin{align}\label{eq:muX}
\mu_{X}=\left|1+\frac{T_{X}^\mathrm{NP}}{T_{X}^\mathrm{SM}}\right|^2\,.
\end{align}
Utilizing the Feynman rules for vertices and propagators outlined in Figure~\ref{fig:feynrules} of Appendix~\ref{app:La}, we can determine the one-loop Feynman diagrams for the decays $h\to Z\gamma$ and $h\to\gamma\gamma$, as depicted in Figures~\ref{fig:htozgamma} and~\ref{fig:htogammagamma}. We have omitted the diagrams containing off-diagonal fermions in the loops, given the strong suppression of the FCNCs. Upon examining the Feynman diagrams for $h\to Z\gamma$ and $h\to\gamma\gamma$, an additional class of diagrams with off-diagonal couplings to the Higgs and $Z$ boson is observed for $h\to Z\gamma$ (i.e., the last row of Figure~\ref{fig:htozgamma}),\footnote{Note that such diagrams have ever appeared in other Higgs models, see, e.g., Refs.~\cite{Bergstrom:1985hp,Gunion:1989we}.} which suggests a greater contribution of the FG2HDM to $h\to Z\gamma$ relative to $h\to\gamma\gamma$. Furthermore, the novel corrections to the $f\bar{f}Z$ couplings offer an alternative pathway to enhance the contribution to $h\to Z\gamma$, as previously discussed in Sec.~\ref{sec:intro}. The one-loop functions $T_{X}$ can then be extracted by matching onto the calculations of the one-loop Feynman diagrams. In this work, we calculate these diagrams using \textit{Package-X}~\cite{Patel:2015tea}, as follows:

\subsection{$h\to Z\gamma$}
\label{subsec:htoZgamma}

\begin{figure}[h!]
  \centering
  \includegraphics[width=0.55\textwidth]{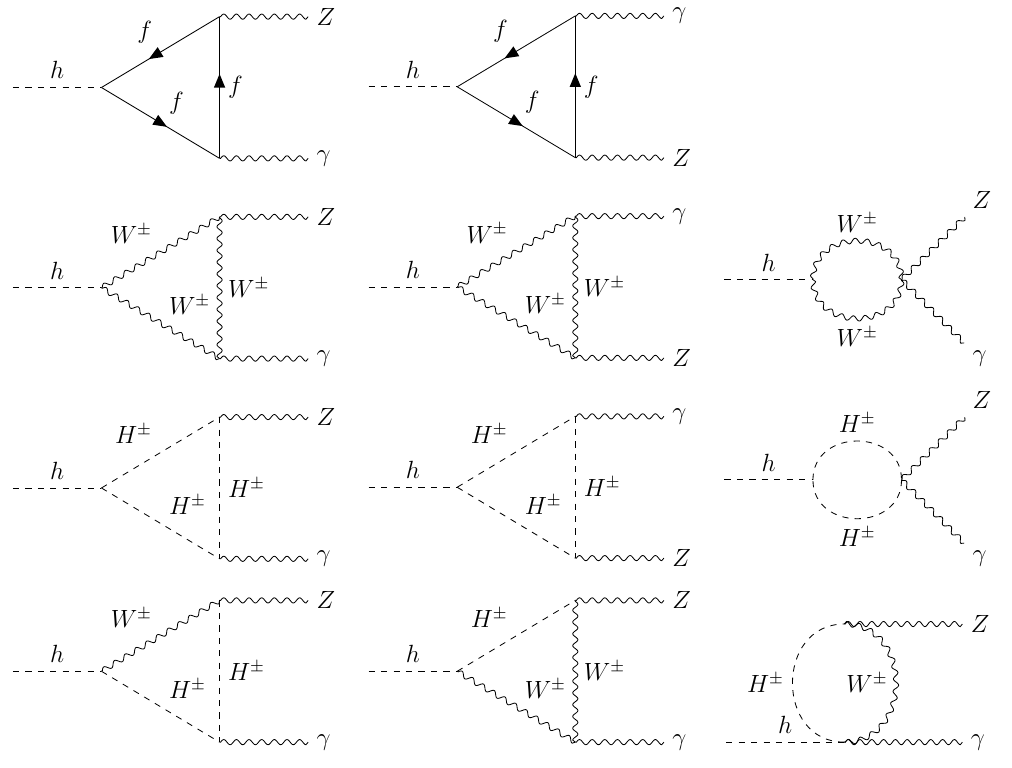}
  \vspace{-0.25cm}
  \caption{\small One-loop Feynman diagrams for $h\to Z\gamma$.}\label{fig:htozgamma}
\end{figure}

The one-loop Feynman diagrams that contribute to the $h\to Z\gamma$ process are depicted in Figure~\ref{fig:htozgamma}. 
We have intentionally omitted the external legs self-energy diagrams for this decay mode because they lack the Lorentz structure specified in Eq.~\eqref{eq:Lorentz}, and the divergences present in the calculations of the Feynman diagrams in each row of Figure~\ref{fig:htozgamma} are found to cancel each other out (a pattern that also applies to $h\to\gamma\gamma$). These diagrams are categorized into four distinct classes (one row, one class), each corresponding to a type of particle involved in the internal lines. Consequently, the total one-loop function $T_{Z\gamma}$ is decomposed into the following four distinct components:
\begin{align}
T_{Z\gamma}=T_{Z\gamma}^f+T_{Z\gamma}^W+T_{Z\gamma}^H+T_{Z\gamma}^{WH}\,,
\end{align}
corresponding to the contributions from fermions, $W$ boson, charged Higgs, and mixed $W$ boson and charged Higgs loops, respectively. Specifically,
\begin{align}
T_{Z\gamma}^{f}=&\sum_f\frac{eg_2Q_f N_Cm_f}{16\pi^2m_W(m_h^2-m_Z^2)^2}\left[\frac{g_2}{c_W}(I_f^3-2Q_fs_W^2)\cos\theta'_2+g'(\mathcal{Q}_{fL}+\mathcal{Q}_{fR})\sin\theta'_2\right]\notag\\\times&\left[\sin(\alpha-\beta)M_f-\cos(\alpha-\beta)N_f\right]\Bigg\{4m_Z^2\left[\Lambda(m_h^2;m_f,m_f)-\Lambda(m_Z^2;m_f,m_f)\right]\notag\\
+&2(m_h^2-m_Z^2)\Big[2+(4m_f^2-m_h^2+m_Z^2)C_0(m_Z^2,0,m_h^2,m_f^2,m_f^2,m_f^2)\Big]\Bigg\}\,,\label{eq:Zgammaf}\\[0.2cm]
T_{Z\gamma}^{W}=&-\frac{eg_2^2c_W \sin(\alpha-\beta)\cos\theta'_2}{16\pi^2m_W^3(m_h^2-m_Z^2)^2}\Bigg\{\left[12m_W^4-2m_W^2m_Z^2+m_h^2(2m_W^2-m_Z^2)\right]\notag\\
\times&\Big[(m_h^2-m_Z^2)+m_Z^2\left(\Lambda(m_h^2;m_W,m_W)-\Lambda(m_Z^2;m_W,m_W)\right)\Big]+2m_W^2(m_h^2-m_Z^2)\notag\\
\times&(12m_W^4\!+\!6m_W^2m_Z^2\!-\!2m_Z^4\!-\!6m_h^2m_W^2\!+\!m_h^2m_Z^2)C_0(m_Z^2,0,m_h^2,m_W^2,m_W^2,m_W^2)\Bigg\}\,,\label{eq:ZgammaW}\\[0.2cm]
T_{Z\gamma}^{H}=&-\frac{4ev\lambda_{hH^+H^-}}{16\pi^2(m_h^2-m_Z^2)^2}\Big[(\frac{g_2}{2}c_W-\frac{g_1}{2}s_W)\cos\theta'_2+(\sin^2\beta\mathcal{Q}_1+\cos^2\beta\mathcal{Q}_2)g'\sin\theta'_2\Big]\notag\\\times&\Bigg\{m_Z^2\Big[\Lambda(m_h^2;m_{H^\pm},m_{H^\pm})-\Lambda(m_Z^2;m_{H^\pm},m_{H^\pm})\Big]\notag\\
+&(m_h^2-m_Z^2)\Big[1+2m_{H^\pm}^2C_0(m_Z^2,0,m_h^2,m_{H^\pm}^2,m_{H^\pm}^2,m_{H^\pm}^2)\Big]\Bigg\}\,,\label{eq:ZgammaH}\\[0.2cm]
T_{Z\gamma}^{WH}=&\frac{eg_2g'(\mathcal{Q}_2-\mathcal{Q}_1)\cos(\alpha-\beta)\sin\beta \cos\beta\sin\theta'_2}{16\pi^2m_W(m_h^2-m_Z^2)}\notag\\
\times&\Bigg\{2(m_h^2-m_{H^\pm}^2+m_W^2)\Bigg[\frac{m_Z^2}{m_h^2-m_Z^2}\left(\Lambda(m_h^2;m_{H^\pm},m_W)-\Lambda(m_Z^2;m_{H^\pm},m_W)\right)\notag\\
+&\frac{m_{H^\pm}^2-m_W^2}{2m_h^2}\log\left(\frac{m_{H^\pm}^2}{m_W^2}\right)+m_{H^\pm}^2 C_0(0,m_h^2,m_Z^2,m_{H^\pm}^2,m_{H^\pm}^2,m_W^2)+1\Bigg]\notag\\
-&2m_W^2(m_h^2+m_{H^\pm}^2-m_W^2-2m_Z^2)C_0(0,m_h^2,m_Z^2,m_W^2,m_W^2,m_{H^\pm}^2)\Bigg\}\,.\label{eq:ZgammaWH}
\end{align}
In Eq.~\eqref{eq:Zgammaf}, we have summed over the contributions from all the fermions, and the color factor $N_C$ is 3 for quarks and 1 for leptons. The symbols $\mathcal{Q}_{Lf}$, $\mathcal{Q}_{fR}$, $\mathcal{Q}_{1}$, and $\mathcal{Q}_{2}$ denote the $U(1)'$ charge of the left-handed fermions, right-handed fermions, $\Phi_{1}$, and $\Phi_{2}$, respectively. The definitions of $M_f$, $N_f$ (Eq.~\eqref{eq:Zgammaf}) and the coupling, $\lambda_{hH^+H^-}$ (Eq.~\eqref{eq:ZgammaH}), are provided in Appendix~\ref{app:La}. The analytical expressions for the scalar functions $\Lambda(m_a^2;m_b,m_c)$, $C_0(m_Z^2,0,m_h^2,m_i^2,m_i^2,m_i^2)$, and $C_0(0,m_h^2,m_Z^2,m_i^2,m_i^2,m_j^2)$ in Eqs.~\eqref{eq:Zgammaf}-\eqref{eq:ZgammaWH} are listed in Appendix~\ref{app:scalarfunc}. Note that similar expressions for $T_X$ can also be found in Ref.~\cite{Fontes:2014xva}. Notably, the novel contribution $T^{WH}_{Z \gamma}$, which arises from the $W^\pm H^\mp Z$ coupling and is uniquely present in FG2HDM, distinguishes it from other conventional 2HDMs.

\subsection{$h\to\gamma\gamma$}
\label{subsec:hto2gamma}

\begin{figure}[h!]
  \centering
  \includegraphics[width=0.55\textwidth]{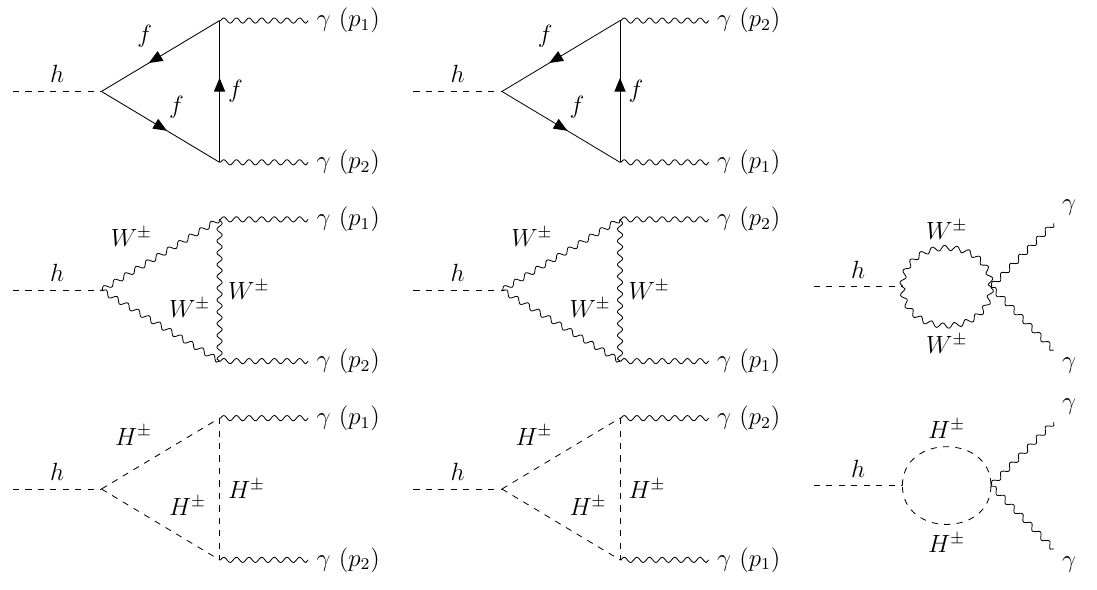}
  \vspace{-0.25cm}
  \caption{\small One-loop Feynman diagrams for $h\to\gamma\gamma$.}\label{fig:htogammagamma}
\end{figure}

The one-loop Feynman diagrams responsible for the $h\to\gamma\gamma$ process are illustrated in Figure~\ref{fig:htogammagamma}. 
In the depiction of the first and second diagrams in each row, distinct momenta $p_1$ and $p_2$ are assigned to differentiate the two identical photons in the final state. 
Paralleling the approach for $h\to Z\gamma$, the total one-loop function $T_{\gamma\gamma}$ is decomposed into three distinct parts:
\begin{align}
T_{\gamma\gamma}=T_{\gamma\gamma}^f+T_{\gamma\gamma}^W+T_{\gamma\gamma}^H\,,
\end{align}
with each term explicitly given by:
\begin{align}
T_{\gamma\gamma}^{f}=&\sum_f\frac{4e^2 g_2 N_C Q_f^2 m_f(\sin(\alpha-\beta)M_f-\cos(\alpha-\beta)N_f)}{16\pi^2m_h^2m_W}\notag\\
\times&\left[2+(4m_f^2-m_h^2)C_0(0,0,m_h^2,m_f^2,m_f^2,m_f^2)\right]\,,\label{eq:2gammaf}\\[0.2cm]
T_{\gamma\gamma}^{W}=&\frac{-2e^2 g_2\sin(\alpha\!-\!\beta)}{16\pi^2m_h^2m_W}\left[(m_h^2\!+\!6m_W^2\!+\!(12m_W^4\!-\!6m_h^2m_W^2)C_0(0,0,m_h^2,m_W^2,m_W^2,m_W^2)\right],\label{eq:2gammaW}\\[0.2cm]
T_{\gamma\gamma}^{H}=&-\frac{4e^2v\lambda_{hH^+H^-}}{16\pi^2m_h^2}\left[1+2m_{H^\pm}^2C_0(0,0,m_h^2,m_{H^\pm}^2,m_{H^\pm}^2,m_{H^\pm}^2)\right]\,.\label{eq:2gammaH}
\end{align}
The analytical expressions for the scalar function $C_0(0,0,m_h^2,m_i^2,m_i^2,m_i^2)$ in Eqs.~\eqref{eq:2gammaf}-\eqref{eq:2gammaH} can also be found in Appendix~\ref{app:scalarfunc}. Unlike in the $h \to Z \gamma$ process, no $T^{WH}_{Z \gamma}$ contribution occurs because the $W^\pm H^\mp\gamma$ interaction is forbidden in FG2HDM, as it is in other conventional 2HDMs and the SM.

\section{Numerical analysis}
\label{sec:num}

In this section, we perform a numerical analysis to assess whether the parameter spaces of FG2HDM can simultaneously accommodate the current measured $\mu_{Z\gamma}$ and $\mu_{\gamma\gamma}$. For ease of reference, the input parameters used in the numerical analysis throughout this work are summarized in Table~\ref{tab:input}. It should be noted that the fine structure constant, $\alpha\equiv e^2/4\pi$, is scale dependent and is related to the electroweak gauge coupling $g_2$ via the relation $e=g_2\sin\theta_W$. Additionally, we do not adopt the experimental value for $\Gamma_h$ from the particle data group (PDG), which has a relatively large uncertainty ($\Gamma_h^\mathrm{PDG}=3.7_{-1.4}^{+1.9}$~MeV)~\cite{ParticleDataGroup:2024cfk}, but rather, we cite the value with smaller uncertainty from Ref.~\cite{LHCHiggsCrossSectionWorkingGroup:2016ypw}. Our strategy in this section initially involves calculating the SM LO contributions for the decays $h\to Z\gamma$ and $h\to\gamma\gamma$, whose numerical values provide valuable insights for NP model buildings. We then incorporate the corrections from the FG2HDM in details. 
\begin{table}[h!]
\centering
\tabcolsep 0.25in
\begin{tabular}{|l|l|}
\hline
$\sin^2\theta_W=0.23129(4)$  \hfill\cite{ParticleDataGroup:2024cfk}  &
$m_W = 80.3692(133)~\mathrm{GeV}$  \hfill\cite{ParticleDataGroup:2024cfk} \\
$\alpha(0)=1/137.036$  \hfill  \cite{ParticleDataGroup:2024cfk} &
$m_Z=91.1880(20)~\mathrm{GeV}$ \hfill\cite{ParticleDataGroup:2024cfk} \\
$\alpha(m_Z)= 1/127.944(14)$  \hfill \cite{ParticleDataGroup:2024cfk} &
$m_h=125.20(11)~\mathrm{GeV}$ \hfill\cite{ParticleDataGroup:2024cfk} \\
$\Gamma_h=4.07^{+4.0\%}_{-3.9\%}~\mathrm{MeV}$ \hfill \cite{LHCHiggsCrossSectionWorkingGroup:2016ypw} & $m_t=172.57(29)~\mathrm{GeV}$ \hfill\cite{ParticleDataGroup:2024cfk} \\
$G_F=1.1663788\times10^{-5}~\mathrm{GeV}^{-2}$ \hfill\cite{ParticleDataGroup:2024cfk} &
$m_{\tau} = 1.77693(9)~\mathrm{GeV}$ \hfill\cite{ParticleDataGroup:2024cfk} \\
$m_b=4.183(7)~\mathrm{GeV}$ \hfill\cite{ParticleDataGroup:2024cfk} & $m_c=1.2730(46)~\mathrm{GeV}$ \hfill\cite{ParticleDataGroup:2024cfk}\\
$m_s=93.5(8)~\mathrm{MeV}$ \hfill\cite{ParticleDataGroup:2024cfk} & $m_\mu=105.658~\mathrm{MeV}$ \hfill\cite{ParticleDataGroup:2024cfk}\\
$m_d=4.70(7)~\mathrm{MeV}$ \hfill\cite{ParticleDataGroup:2024cfk} & $m_e=0.511~\mathrm{MeV}$ \hfill\cite{ParticleDataGroup:2024cfk}\\
$m_u=2.16(7)~\mathrm{MeV}$ \hfill\cite{ParticleDataGroup:2024cfk} & \\
\hline
\end{tabular}
\caption{\small Relevant input parameters used in our numerical analysis.}
\label{tab:input}
\end{table} 

\subsection{SM prediction}

The expressions in Eqs.~\eqref{eq:Zgammaf}-\eqref{eq:ZgammaWH} and~\eqref{eq:2gammaf}-\eqref{eq:2gammaH} revert to the SM formulas upon setting $\theta'_2\to 0$ and $\sin(\alpha-\beta)\to 1$. The numerical results for $T_{Z\gamma}$ (in the unit of GeV$^{-1}$) in the SM are then given by:
\begin{align}\label{eq:SMTZgamma}
T_{Z\gamma}^{W,\text{SM}}=&-5.866\times10^{-5}\,, & T_{Z\gamma}^{t,\text{SM}}=&3.115\times10^{-6}\,,\notag\\[0.2cm]
T_{Z\gamma}^{b,\text{SM}}=&-6.685\times10^{-8}+3.762\times10^{-8}i\,, & T_{Z\gamma}^{c,\text{SM}}=&-9.797\times10^{-9}+3.864\times10^{-9}i\,,\notag\\[0.2cm]
T_{Z\gamma}^{\tau,\text{SM}}=&-1.731\times10^{-9}+7.452\times10^{-10}i\,. & &
\end{align}
Eq.~\eqref{eq:SMTZgamma} (and also Eq.~\eqref{eq:SMT2gamma}) clearly indicates that the dominant contributions arise from the $W$ loops, and the secondary contributions come from the interference between the $W$ and top quark loops. Contributions from the $s$, $d$, $u$, $\mu$, and $e$ loops are not presented here, as they are several orders of magnitude smaller than those we have displayed and therefore can be neglected safely. Likewise, the SM contributions to $T_{\gamma\gamma}$ (in the unit of GeV$^{-1}$) are
\begin{align}\label{eq:SMT2gamma}
T_{\gamma\gamma}^{W,\text{SM}}=&-3.912\times10^{-5}\,, & T_{\gamma\gamma}^{t,\text{SM}}=&8.619\times10^{-6}\,,\notag\\[0.2cm]
T_{\gamma\gamma}^{b,\text{SM}}=&-1.123\times10^{-7}+1.485\times10^{-7}i\,, & T_{\gamma\gamma}^{c,\text{SM}}=&-9.103\times10^{-8}+7.462\times10^{-8}i\,,\notag\\[0.2cm]
T_{\gamma\gamma}^{\tau,\text{SM}}=&-1.107\times10^{-7}+1.011\times10^{-7}i\,. & &
\end{align}
With the kinematics formula derived in Appendix~\ref{app:kin}, the SM predictions for $h\to Z\gamma$ and $h\to\gamma\gamma$ are given separately by
\begin{align}
\mathcal{B}(h\to Z\gamma)_\mathrm{SM}=&(1.536\pm0.018)\times10^{-3}\,,\\[0.2cm]
\mathcal{B}(h\to\gamma\gamma)_\mathrm{SM}=&(2.278\pm0.023)\times10^{-3}\,,
\end{align}
which are respectively in good agreements with those given in Ref.~\cite{LHCHiggsCrossSectionWorkingGroup:2016ypw}: $\mathcal{B}(h\to Z\gamma)=(1.5\pm0.1)\times10^{-3}$, and $\mathcal{B}(h\to\gamma\gamma)=(2.27\pm0.07)\times10^{-3}$.
 
\subsection{Constraining FG2HDM parameters}
\label{subsec:FG2HDM}

The $h\to\gamma\gamma$ decay imposes stringent constraints on the FG2HDM parameters, as the SM prediction of $\mu_{\gamma\gamma}$ aligns well with its measurement. This implies that we should have $\sin(\alpha-\beta)\approx 1$ (or equivalently, $\cos(\alpha-\beta)\approx 0$) to ensure that $T_{\gamma\gamma}^f$ and $T_{\gamma\gamma}^W$ are close to their SM values, and the contribution from the charged Higgs loops should be minimal. Additionally, we should have $\cos\theta'_2\approx 1$ (or equivalently, $\sin\theta'_2\approx 0$) to ensure that the $ZW^+W^-$ coupling does not deviate too much from its SM value. 

With the aforementioned approximations, we can first determine the parameter regions for the charged Higgs interactions. They can be classified into two parts, i.e., the pure charged Higgs diagrams and the $W^\pm$-$H^\pm$ mixing diagrams. Note that the latter only contribute to $h\to Z\gamma$, which is actually negligible compared to the pure one. The analysis is as follows.

\begin{figure}[htb!]
  \centering
  \includegraphics[width=0.485\textwidth]{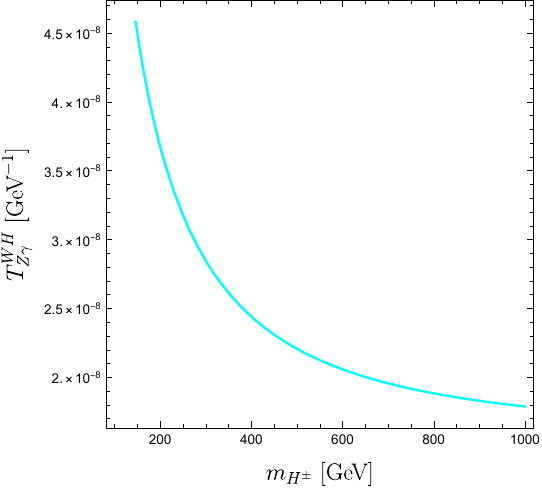}
  \includegraphics[width=0.46\textwidth]{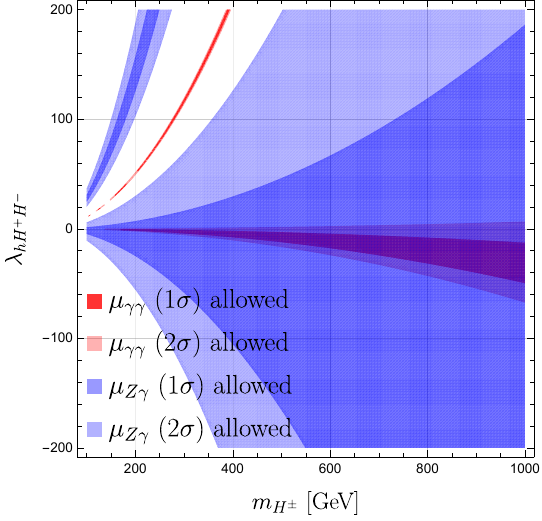}
  \vspace{-0.25cm}
  \caption{\small Left: Magnitude of $T_{Z\gamma}^{WH}$ as a function of $m_{H^\pm}$ with input $|a|=0.01$. Right: the $1\sigma$ and $2\sigma$ allowed regions for $m_{H^\pm}$ and $\lambda_{hH^+H^-}$ with constraints from $\mu_{Z\gamma}$ (blue) and $\mu_{\gamma\gamma}$ (red).}\label{fig:chargedHiggs}
\end{figure}

To estimate the magnitude of $T_{Z\gamma}^{WH}$ given in Eq.~\eqref{eq:ZgammaWH}, we need to endow values for $\cos(\alpha-\beta)$ (or equivalently, $\sin(\alpha-\beta)$), $g'$, $\mathcal{Q}_2-\mathcal{Q}_1$, $\cos\beta$ (or equivalently, $\sin\beta$), and $\sin\theta'_2$. However, as we do not have the exact values for these parameters, we can only make some rough but reasonable estimates. As discussed in the first paragraph of this subsection, to ensure that $T_{\gamma\gamma}^f$ and $T_{\gamma\gamma}^W$ do not deviate too much from their SM predictions, we should have $\cos(\alpha-\beta)\approx 0$ and $\sin\theta'_2\approx 0$. Note that the constraints from the $B$ observables in our previous work suggest that $\tan\beta<28$ and $g'>10^{-2}$~\cite{Wen:2025smu}. The large upper limit of $\tan\beta$ suggests that $\cos\beta$ should be small, whereas $g'$ is expected to remain sufficiently small to preserve the perturbative nature of the $U(1)'$ gauge theory. Furthermore, it is reasonable to assume that $\mathcal{Q}_2-\mathcal{Q}_1$ is of $\mathcal{O}(1)$ under the consideration of naturalness. While satisfying these conditions and simultaneously hoping that $T_{Z\gamma}^{WH}$ is as large as possible, we assume that $\cos(\alpha-\beta)\sim\cos\beta\sim \sin\theta'_2 \sim \mathcal{O}(10^{-1})$ and $g'\sim(\mathcal{Q}_2-\mathcal{Q}_1)\sim\mathcal{O}(1)$. With these assumptions, we can conservatively estimate:
\begin{align}
a\equiv g'(\mathcal{Q}_2-\mathcal{Q}_1)\cos(\alpha-\beta)\sin\beta \cos\beta\sin\theta'_2\sim\mathcal{O}(10^{-3})\,.
\end{align}
To make a maximum estimate on $T_{Z\gamma}^{WH}$, let us assume that $|a|$ is close to the margin of $\mathcal{O}(10^{-2})$, i.e., $|a|=0.01$. With this entry, we plot the magnitude of $T_{Z\gamma}^{WH}$ as a function of $m_{H^\pm}$ in the left panel of Figure~\ref{fig:chargedHiggs}. Obviously, $T_{Z\gamma}^{WH}$ decreases as $m_{H^\pm}$ increases: the magnitude of which can only reach $\mathcal{O}(10^{-8})$ GeV$^{-1}$, comparable to that of the $b$ quark loops in the SM (cf. the value of $T_{Z\gamma}^{b,\text{SM}}$ in  Eq.~\eqref{eq:SMTZgamma}), which is one and two orders of magnitude smaller than that of the $t$ quark loops and $W$ loops,  respectively. Therefore, the contribution from the $W^\pm$-$H^\pm$ mixing diagrams is negligible. Below, we merely consider the contributions from the pure charged Higgs diagrams.

In the right panel of Fig.~\ref{fig:chargedHiggs}, we present the $1\sigma$ and $2\sigma$ allowed regions for the parameters $m_{H^\pm}$ and $\lambda_{hH^+H^-}$, with constraints derived from $\mu_{Z\gamma}$ (blue) and $\mu_{\gamma\gamma}$ (red). It is clear from this plot that in the parameter regions where $m_{H^\pm}>200$~GeV and $\lambda_{hH^+H^-}<0$, both the measured $\mu_{Z\gamma}$ and $\mu_{\gamma\gamma}$ can be accommodated within the $1\sigma$ level. Interestingly, the bound on the charged Higgs mass obtained in this work is also consistent with that from the constraints of $B$ observables in Ref.~\cite{Wen:2025smu} (see the right panel of Fig.~1 in this reference). As discussed above, since $\mu_{\gamma\gamma}$ is better than $\mu_{Z\gamma}$ in terms of the consistency between the SM prediction and measurement, the overlapped parameter regions allowed by the two observables in the right panel of Figure~\ref{fig:chargedHiggs} are mainly narrowed down by the former.

The remaining contribution comes from the fermion loop diagrams, in which the FG2HDM influence the $h\to Z\gamma$ decay by correcting the $f\bar{f}Z$ couplings. As shown in the Feynman diagrams depicted in Figs~\ref{fig:htozgamma} and~\ref{fig:htogammagamma}, such corrections only influence $\mu_{Z\gamma}$ and leave $\mu_{\gamma\gamma}$ entirely unaffected, as the latter does not involve any $f\bar{f}Z$ vertex. Although the $f\bar{f}Z$ vertex corrections also influence the $h\to ZZ^\ast$ decay, as two $f\bar{f}Z$ vertices are involved in such a process, the impacts on them is negligible compared with that on the $h\to Z\gamma$ decay, as the latter contains only one $f\bar{f}Z$ vertex. Therefore, we do not invoke $h\to ZZ^\ast$ in the numerical constraints.

Note that the total $f\bar{f}Z$ coupling in Eq.~\eqref{eq:FG} has been projected into the left-handed part $C_L^f$ and right-handed part $C_R^f$, with 
\begin{align}
C_L^f=\frac{g_2\cos\theta'_2}{c_W}(I_f^3-Q_fs_W^2)+g'\sin\theta'_2\mathcal{Q}_{fL}\,,\qquad
C_R^f=-\frac{g_2\cos\theta'_2}{c_W}Q_fs_W^2+g'\sin\theta'_2\mathcal{Q}_{fR}\,.
\end{align}
In the limit $\cos\theta'_2\sim 1$ ($\sin\theta'_2\sim 0$), the above expressions restore to the ones in the SM. Given the dominant role of the top quark in contributions from fermion loops, particular emphasis is placed on the correction from the $t\bar{t}Z$ vertex. Using $I_t^3=1/2$ and $Q_t=+2/3$ for the top quark, along with the values of $s_W$ and $\alpha(m_Z)$ from Table~\ref{tab:input}, we obtain $C_L^{t,\mathrm{SM}}\simeq0.256$ and $C_R^{t,\mathrm{SM}}\simeq-0.114$. Notice that in Eq.~\eqref{eq:Zgammaf} the contribution from the top loops depends on the sum of $C_L^t$ and $C_R^t$, i.e., the vector coupling $C_V^t\equiv(C_L^t+C_R^t)/2$. Therefore, 
the deviation $\Delta C_L^t\equiv(\Delta C_L^t+\Delta C_R^t)/2$, where 
\begin{align}
\Delta C_L^t=g'\sin\theta'_2\mathcal{Q}_{tL}\,,\qquad \Delta C_R^t=g'\sin\theta'_2\mathcal{Q}_{tR}\,,
\end{align}
measures the deviation from the SM, i.e., the contribution from FG2HDM. The measurement of $\mu_{Z\gamma}$ can impose a constraint on $\Delta C_V^t$, which subsequently can be translated into the bounds on the parameters of FG2HDM. 

Besides $\mu_{Z\gamma}$, $\Delta C_V^t$ also subjects to constraints from the following two primary sources: i) top quark observables, and ii) the FCNC process $b\to s\ell^+\ell^-$. Regarding the former, both the CMS~\cite{CMS:2019too} and ATLAS~\cite{ATLAS:2021fzm,ATLAS:2023eld} collaborations have conducted precise measurements of the differential and/or inclusive cross section of top quark pair production in association with a $Z$ boson at the LHC. ATLAS has also measured the production of single top quark and anti-top quark via the $t$-channel exchange of a $W$ boson~\cite{ATLAS:2024ojr}. These measurements align well with their SM predictions and can be leveraged to establish limits on the Wilson coefficients of the SM effective field theory (SMEFT)~\cite{Buchmuller:1985jz,Grzadkowski:2010es,Brivio:2017vri}. By invoking the matching between $\Delta C_V^t$ and the SMEFT Wilson coefficients~\cite{Aguilar-Saavedra:2008nuh}:
\begin{align}\label{eq:match}
\Delta C_V^t=\frac{g_2v^2}{2\Lambda^2c_W}\mathrm{Re}\left[-C_{H u}^{33}-C_{H Q}^{(1,33)}+C_{H Q}^{(3,33)}\right]\,,
\end{align}
where $C_{H u}$, $C_{H Q}^{(1)}$, and $C_{H Q}^{(3)}$ denote respectively the Wilson coefficients of the following SMEFT operators,
\begin{align}
\mathcal{O}_{H u}\equiv&(H^\dagger i\overleftrightarrow{D}_\mu H)(\bar{u}\gamma^\mu u)\,,\notag\\
\mathcal{O}_{H Q}^{(1)}\equiv&(H^\dagger i\overleftrightarrow{D}_\mu H)(\bar{Q}\gamma^\mu Q)\,,\notag\\
\mathcal{O}_{H Q}^{(3)}\equiv&(H^\dagger i\overleftrightarrow{D}_\mu\tau^I H)(\bar{Q}\gamma^\mu\tau^I Q)\,,
\end{align}
we can delineate the permissible range for $\Delta C_V^t$. Adopting the $95\%$ CL intervals provided by ATLAS~\cite{ATLAS:2023eld}: 
\begin{align}
C_{H t}^{33}\in[-2.2,1.6]\,,\qquad C_{H Q}^{(1,33)}\in[-1.4,0.84]\,,\qquad C_{H t}^{33}\in[-0.95,2.0]\,,
\end{align}
which are obtained with $\Lambda=1$~TeV and when one coefficient at a time is assumed to be non-zero, we obtain $\Delta C_V^t\in[-0.076,0.126]$. 
Regarding the $b \to s \ell^+ \ell^-$ constraint, the anomalous coupling $\Delta C_V^t$ contributes to this FCNC process via the insertion of the $t\bar{t}Z$ vertex into the $Z$-penguin diagram. This modification affects the Wilson coefficients $C_9$ and $C_{10}$, which are associated with the operators $\mathcal{O}_9 \equiv (\bar{s}\gamma_\mu P_L b)(\bar{\ell}\gamma^\mu\ell)$ and $\mathcal{O}_{10} \equiv (\bar{s}\gamma_\mu P_L b)(\bar{\ell}\gamma^\mu\gamma_5\ell)$ ($P_L = \frac{1-\gamma_5}{2}$), respectively. Beyond these loop-level effects, FG2HDM introduces additional tree-level FCNC amplitudes that contribute to $C_9$ and $C_{10}$. These contributions depend on the same set of model parameters as $\Delta C_V^t$. Consequently, our numerical strategy begins with a comprehensive parameter scan to identify the allowed ranges of the model parameters. These results are then used to determine the viable range for $\Delta C_V^t$. The total corrections to $C_9$ and $C_{10}$ are expressed as~\cite{Wen:2023pfq}
\begin{align}
\Delta C_{9} &= -\frac{1}{N} \left( \frac{1}{m_{Z'}^2}\mathcal{A}_L^{bsZ'}\mathcal{B}^{Z'\ell\ell} + \frac{1}{m_{Z}^2}\mathcal{A}_L^{bsZ}\mathcal{B}^{Z\ell\ell} \right)\,, \\
\Delta C_{10} &= -\frac{1}{N} \left( \frac{1}{m_{Z'}^2}\mathcal{A}_L^{bsZ'}\mathcal{B}_5^{Z'\ell\ell} + \frac{1}{m_{Z}^2}\mathcal{A}_L^{bsZ}\mathcal{B}_5^{Z\ell\ell} \right)\,,
\end{align}
where $N = 2\sqrt{2}G_F V_{tb}V_{ts}^\ast \alpha / (4\pi)$, and $m_Z$ ($m_{Z'}$) denotes the mass of the $Z$ ($Z'$) boson, representing the contributions from $Z$-exchange ($Z'$-exchange) diagrams. The functions $\mathcal{A}_{L}$ and $\mathcal{B}_{(5)}$ depend on the FG2HDM parameter set $\{g', \mathcal{Q}_{\mu_R}, \mathcal{Q}_{d_R}, \sin\theta_2'\}$, for which detailed expressions are provided in Ref.~\cite{Wen:2025smu}. By fixing $\sin\theta_2' = 0.1$, requiring $m_{Z'} < 450$ GeV, and incorporating the established constraints, our scan reveals that $\Delta C_V^t$ is constrained within $[-0.05, 0.05]$ at the $3\sigma$ level.

\begin{figure}[htb!]
  \centering
  \includegraphics[width=0.46\textwidth]{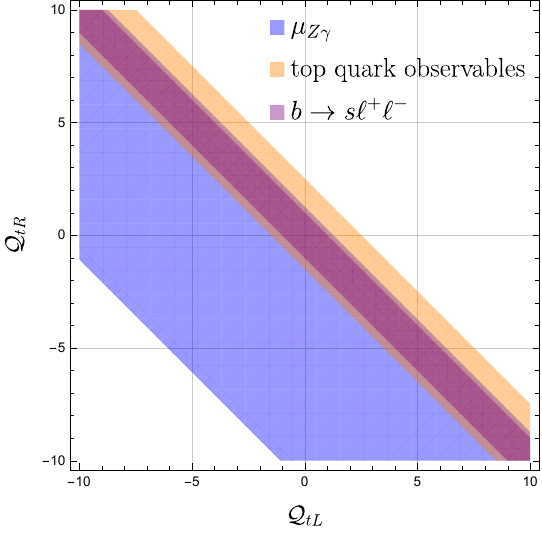}
  \vspace{-0.25cm}
  \caption{\small The $\mathcal{Q}_{tL}$-$\mathcal{Q}_{tR}$ allowed regions obtained from $\mu_{Z\gamma}$ (blue), top quark observables (orange), and $b\to s\ell^+\ell^-$ (purple), with $g'=1$ and $\sin\theta'_2=0.1$.}\label{fig:ttZ}
\end{figure}

Fixing $g'=1$ and $\sin\theta'_2=0.1$,\footnote{The numerical constraints with assigning other values for $g'$ and $\sin\theta'_2$ are similar, and they differ from each other only by a rescale factor.} and designating $\mathcal{Q}_{tL}$ and $\mathcal{Q}_{tR}$ as the two independent parameters that need to be restricted, the constraints on which from $\mu_{Z\gamma}$ (blue), top quark observables (orange), and $b\to s\ell^+\ell^-$ (purple) are shown in Figure~\ref{fig:ttZ}. It is clear from this plot that there is an overlapped region for $\mathcal{Q}_{tL}$ and $\mathcal{Q}_{tR}$ to simultaneously reconcile the three observables. However, due to the relatively large uncertainty, the constraint from $\mu_{Z\gamma}$ is weaker than that from top quark observables and $b\to s\ell^+\ell^-$. It is also observed that till now the most stringent bound in this respect is taken from the low-energy process $b\to s\ell^+\ell^-$ decay. 

By utilizing the parameters derived above, we evaluate the contributions of different $T$ functions to $\mu_{Z\gamma}$ and $\mu_{\gamma\gamma}$. The results show that $T_{Z\gamma}^W$ and $T_{\gamma\gamma}^W$ are of the order $10^{-5}$ GeV$^{-1}$, consistent with the SM, whereas $T_{Z\gamma}^H$, $T_{\gamma\gamma}^H$, and $T_{Z\gamma}^t$ are around $10^{-6}$ GeV$^{-1}$. Due to this clear hierarchy, the $W$-boson contributions dominate the amplitudes. Interference terms involving $W$ are sub-dominant, and all remaining cross-terms are effectively negligible. This justifies our approach of analysing the constraints from each $T$ function separately rather than accounting for their combined interference effects.

\section{Conclusion}
\label{sec:con}

In this paper, we investigated the $h\to Z\gamma$ and $h\to \gamma\gamma$ decays within the framework of FG2HDM. FG2HDM is a BSM model based on the $SU(3)_C\times SU(2)_L\times U(1)_Y\times U(1)'$ gauge symmetry, extending the SM particle spectrum with five additional physical scalars and one neutral gauge boson, $Z'$. We derived the Feynman rules for the model and calculated the one-loop amplitudes for both decay channels. We found that while both channels are influenced by the charged Higgs bosons, $h \to Z\gamma$ is uniquely modified by corrections to the $f\bar{f}Z$ vertex. These vertex corrections are also subject to constraints from top-quark observables and the FCNC process $b\to s\ell^+\ell^-$. Our numerical analysis demonstrated that the amplitude contributions from $W^\pm$-$H^\pm$ mixing diagrams are comparable to the SM $b$-quark loop contributions, which are known to be negligible. Consequently, we focused our analysis on the contributions from pure charged Higgs diagrams. We showed that a parameter region with $m_{H^\pm}>200$~GeV and $\lambda_{hH^+H^-}<0$ can simultaneously accommodate both $\mu_{Z\gamma}$ and $\mu_{\gamma\gamma}$ measurements at the $1\sigma$ level. This region is primarily constrained by the latter owing to its relatively small uncertainty. Regarding the constraints on the $f\bar{f}Z$ vertex corrections, we considered only the dominant contribution from the top quark. By fixing $g'=1$ and $\sin\theta'_2=0.1$, we identified a viable region that simultaneously satisfies the constraints from $\mu_{Z\gamma}$, top-quark observables, and $b\to s\ell^+\ell^-$, with the strongest constraints coming from $b\to s\ell^+\ell^-$. In summary, owing to the currently large uncertainty associated with $\mu_{Z\gamma}$, its constraining power is limited compared to other precision observables. Nevertheless, as discussed in the previous section, the HL-LHC with a projected integrated luminosity of $3~\rm{ab}^{-1}$ is expected to reduce the relative uncertainty of $\mu_{Z\gamma}$ to $14\%$~\cite{ATLAS:2025eii}. This improvement would significantly strengthen the constraints on FG2HDM.

\textbf{Note Added.} All the authors contribute equally and they are co-first authors, while F. Xu is the corresponding author.

\section*{Acknowledgements}

This work is supported by NSFC under Grant Nos.~12475095 and~U1932104, the Fundamental Research Funds for the Central Universities (11623330), and the 2024 Guangzhou Basic and Applied Basic Research Scheme Project for Maiden Voyage (2024A04J4190).

\appendix

\section{Lagrangian and Feynman rules}
\label{app:La}

Since at least one photon is involved in both the $h\to Z\gamma$ and $h\to\gamma\gamma$ decays, the propagators of the one-loop diagrams shall always be charged particles (see Figs~\ref{fig:htozgamma} and~\ref{fig:htogammagamma}). Therefore, to calculate the amplitudes we should first identify the Lagrangian that describe the interactions between the charged particles and the Higgs, $Z$ boson, and photon. The formalisms of Feynman rules depend on the choice of gauge. In this paper, we focus on the unitary gauge, in which the Goldstone bosons disappear and only physical particles participate the interactions. After scrutinizing the total Lagrangian provided in Ref.~\cite{Wen:2025smu}, we list the most relevant terms to calculate the amplitudes of $h\to Z\gamma$ and $h\to\gamma\gamma$ decays as follows:
\begin{align}
\mathcal{L}_{S}=&-\lambda_{hH^+H^-}vhH^+H^-\,,\label{eq:S}\\
\mathcal{L}_{SF}=&-\frac{1}{v}\left[\sin(\alpha-\beta)M_f-\cos(\alpha-\beta)N_f\right]h\bar{f}f\,,\label{eq:SF}\\
\mathcal{L}_{SG}=&\frac{g_2^2v}{2}\sin(\alpha-\beta)hW_\mu^+W^{-\mu}
-ie A_\mu(\partial^\mu H^+H^--\partial^\mu H^-H^+)+e^2A_\mu A^\mu H^+H^-\notag\\
-&i\Big[(\frac{g_2}{2}c_W-\frac{g_1}{2}s_W)\cos\theta'_2+(\sin^2\beta\mathcal{Q}_1+\cos^2\beta\mathcal{Q}_2)g'\sin\theta'_2\Big]Z_\mu(\partial^\mu H^+H^--\partial^\mu H^-H^+)\notag\\
-&i\frac{g_2^2}{2}\cos(\alpha-\beta)\left[\partial^\mu h(W_\mu^+H^--W_\mu^-H^+)+h(\partial^\mu H^+W_\mu^--\partial^\mu H^-W_\mu^+)\right]\notag\\
+&g_2g'(\mathcal{Q}_2-\mathcal{Q}_1)v\sin\beta\cos\beta\sin\theta'_2Z^\mu(W_\mu^+H^-+W_\mu^-H^+)\notag\\
+&2g_2s_W\Big[(\frac{g_2}{2}c_W-\frac{g_1}{2}s_W)\cos\theta'_2+(\sin^2\beta\mathcal{Q}_1+\cos^2\beta\mathcal{Q}_2)g'\sin\theta'_2\Big]A_\mu Z^\mu H^+H^-\notag\\
-&\frac{g_2^2}{2}s_W\cos(\alpha-\beta)A^\mu (H^+W_\mu^-+H^-W_\mu^+)h\,,\label{eq:SG}\\
\mathcal{L}_{G}=&ig_2c_W\cos\theta_{2}^\prime\big[\left(\partial_\mu W_\nu^+W^{-\mu}Z^\nu-\partial_\mu W_\nu^+W^{-\nu}Z^\mu\right)-\left(\partial_\mu W_\nu^-W^{+\mu}Z^\nu-\partial_\mu W_\nu^-W^{+\nu}Z^\mu\right)\notag\\
+&\left(\partial_\mu Z_\nu W^{+\mu}W^{-\nu}-\partial_\mu Z_\nu W^{-\mu}W^{+\nu}\right)\big]\notag +ig_2s_W\big[\left(\partial_\mu W_{\nu}^+W^{-\mu}A^\nu-\partial_\mu W_\nu^+W^{-\nu}A^\mu\right)\notag\\
-&\left(\partial_\mu W_{\nu}^-W^{+\mu}A^\nu-\partial_\mu W_{\nu}^-W^{+\nu}A^{\mu}\right)
+\left(\partial_\mu A_\nu W^{+\mu}W^{-\nu}-\partial_\mu A_\nu W^{-\mu}W^{+\nu}\right)\big]\notag\\
-&g_2^2s_Wc_W\cos\theta_{2}^\prime\big[2W^+_\mu W^{-\mu}A^\nu Z_\nu-W_\mu^+ W^{-\nu}A_\nu Z^\mu-W_\nu^-W^{+\mu}A_\mu Z^\nu\big]\notag\\
-&g_2^2s_W^2\left(W^+_\mu W^{-\mu}A^\nu A_\nu-W^+_\mu W^{-\nu}A_\nu A^\mu\right)\,,\label{eq:G}\\
\mathcal{L}_{FG}=&\frac{g_2\cos\theta_{2}^\prime}{c_{W}}Z_\mu\bar{f}\left[(I_f^3-Q_fs_{W}^2)\gamma^\mu P_L-Q_fs_{W}^2\gamma^\mu P_R\right]f \notag\\
		+&g'\sin\theta_{2}^\prime Z_\mu\bar{f}\left[\mathcal{Q}_{fL}\gamma^\mu P_L+\mathcal{Q}_{fR}\gamma^\mu P_R\right]f+eQ_fA_\mu\bar{f}\gamma^\mu f\,.\label{eq:FG}
\end{align}
The first Lagrangian $\mathcal{L}_{S}$, which describes the interaction between the neutral and charged Higgs, is obtained by expressing the scalar potential (cf. Eq.~\eqref{eq:potential}) in terms of physical fields and their coupling reads
\begin{align}
\lambda_{hH^+H^-}&=\frac{1}{v}\big[\lambda_{1}v_1\sin\alpha\sin^2\beta-\lambda_{2}v_2\cos\alpha\cos^2\beta+v_1\cos\beta\left(\lambda_{3}\sin\alpha\cos\beta+\lambda_{4}\cos\alpha\sin\beta\right)\notag\\
&-v_2\sin\beta\left(\lambda_{3}\cos\alpha\sin\beta+\lambda_4\sin\alpha\cos\beta\right)\big]\,.
\end{align} 
The second Lagrangian $\mathcal{L}_{SF}$ is the Yukawa interactions, with $M_f$ ($f=u,d,\ell$) being the diagonal mass matrices: $M_u=\text{diag}(m_u,m_c,m_t)$, $M_d=(m_d,m_s,m_b)$, and $M_\ell=(e,\mu,\tau)$. Here, $M_f$ is diagonalized by a biunitary transformation:
\begin{align}
M_f=\frac{1}{\sqrt{2}}U_{fL}^\dagger\left(v_1Y_1^f+v_2Y_2^f\right)U_{fR}\,,
\end{align}
where $Y_1^f$ and $Y_2^f$ are Yukawa matrices, and $U_{fL}$ and $U_{fR}$ are unitary matrices. For convenience, we have also introduced the auxiliary matrices $N_f$, which are defined as
\begin{align}
N_f=\frac{1}{\sqrt{2}}U_{fL}^\dagger\left(v_1Y_2^f-v_2Y_1^f\right)U_{fR}\,.
\end{align}
To investigate the $B$-anomalies reported in Ref.~\cite{Wen:2025smu}, we employ specific $U(1)'$ charge assignments for the fields. This configuration ensures that tree-level FCNCs emerge specifically in the down-type quark sector. With such assignments, the Yukawa matrices have the following textures:
\begin{align}\label{eq:YukawaTexture}
    Y_1^u =& \left( \begin{array}{ccc}
        * & * & 0 \\
        * & * & 0 \\
        0 & 0 & 0 
    \end{array} \right)\,, &
    Y_2^u =& \left( \begin{array}{ccc}
        0 & 0 & 0 \\
        0 & 0 & 0 \\
        0 & 0 & * 
    \end{array} \right)\,, &
    Y_1^d =& \left( \begin{array}{ccc}
        * & * & * \\
        * & * & * \\
        0 & 0 & 0 
    \end{array} \right)\,, &
    Y_2^d =& \left( \begin{array}{ccc}
        0 & 0 & 0 \\
        0 & 0 & 0 \\
        * & * & * 
    \end{array} \right)\,, \notag\\
    Y_1^\ell =& \left( \begin{array}{ccc}
        0 & 0 & 0 \\
        0 & * & 0 \\
        0 & 0 & * 
    \end{array} \right)\,, &
    Y_2^\ell =& \left( \begin{array}{ccc}
        * & 0 & 0 \\
        0 & 0 & 0 \\
        0 & 0 & 0 
    \end{array} \right)\,, 
\end{align}
where `$*$' denotes an arbitrary non-zero number. Then the auxiliary matrices $N_f$ have the following explicit forms:
\begin{align}\label{eq:Nf}
    N_u &= -\frac{v_2}{v_1} \,{\rm diag}(m_u, m_c, 0) + \frac{v_1}{v_2} \, {\rm diag}(0, 0, m_t)\,, \nonumber \\
    (N_d)_{ij} &= -\frac{v_2}{v_1} (M_d)_{ij} + \left(\frac{v_2}{v_1}+\frac{v_1}{v_2}\right) (V^\dagger_{\rm CKM})_{i3} (V_{\rm CKM})_{3j} (M_d)_{jj}\,, \nonumber \\
     N_\ell &= -\frac{v_2}{v_1} \textrm{diag}(0, m_\mu, m_\tau) + \frac{v_1}{v_2} \textrm{diag}(m_e, 0, 0)\,,
\end{align}
where $V_{\rm CKM}$ denotes the Cabibbo-Kobayashi-Maskawa (CKM) matrix~\cite{Cabibbo:1963yz,Kobayashi:1973fv}. The third Lagrangian $\mathcal{L}_{SG}$ describes the interactions between the scalars and the gauge bosons, which is obtained by expanding the scalar kinetic terms with gauge bosons contained in the covariant derivatives. The quantities $\mathcal{Q}_1$ and $\mathcal{Q}_2$ denote respectively the $U(1)'$ charges of $\Phi_1$ and $\Phi_2$, and $s_W$ and $c_W$ are separately short for $\sin\theta_W$ and $\cos\theta_W$. The fourth Lagrangian $\mathcal{L}_{G}$ describes the interactions between the gauge bosons, which results from the vector field strength terms. The last Lagrangian $\mathcal{L}_{FG}$ contains the interactions between the fermions and $Z$ boson and photon, where $I_f^3$ denotes the third component of the weak isospin of a fermion doublet, $Q_f$ denotes the electric charge of a given fermion $f$, and $\mathcal{Q}_{fL}$ and $\mathcal{Q}_{fR}$ represent the $U(1)'$ charge of a left-handed and right-handed fermion, respectively. It is also clear from Eq.~\eqref{eq:FG} that the $Z-Z'$ mixing yields an additional term that has the same Lorentz structure as the SM one, which therefore provides a correction to the original couplings. For more details of FG2HDM Lagrangian in different sectors, refer to Ref.~\cite{Wen:2025smu}.

\begin{figure}[h!]
  \centering
  \includegraphics[scale=0.58]{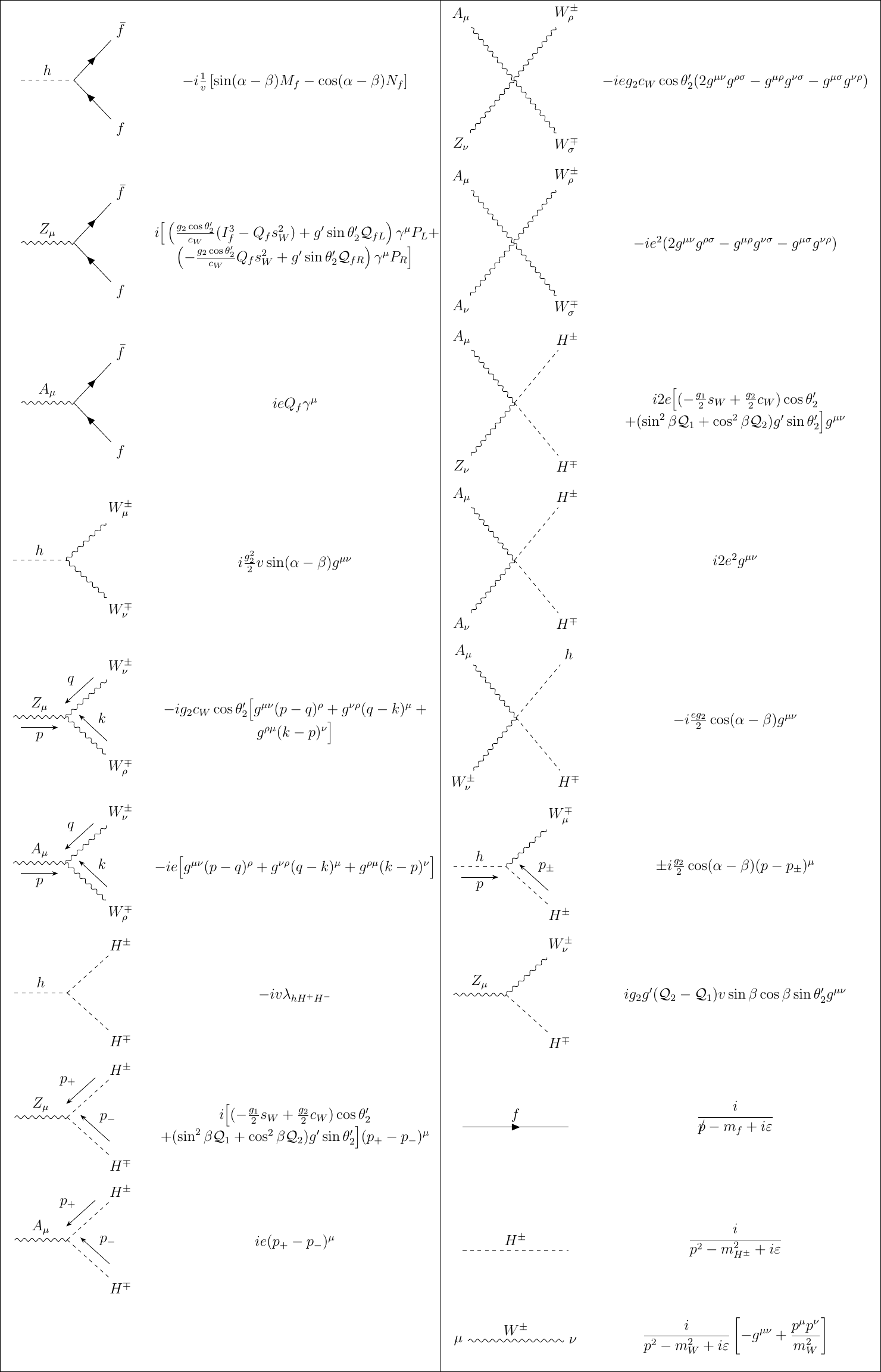}
  \vspace{-0.25cm}
  \caption{\small Feynman rules for the relevant vertices and propagators in the unitary gauge.}\label{fig:feynrules}
\end{figure}

With the Lagrangian listed in Eqs.~\eqref{eq:S}-\eqref{eq:FG} at hand, obtaining the Feynman rules for the vertices is straightforward. Besides, the Feynman rules for the propagators of the charged Higgs, $W$ boson, and fermions can be derived directly from the free Lagrangian that we do not show. Working in the unitary gauge, we summarize all the relevant vertices as well as propagators of FG2HDM that are necessary to calculate the amplitudes of the $h\to Z\gamma$ and $h\to\gamma\gamma$ decays in Fig.~\ref{fig:feynrules}.

\section{Scalar functions}\label{app:scalarfunc}

In this appendix, we show the analytical expressions for the scalar functions appearing in Sec.~\ref{subsec:htoZgamma} and~\ref{subsec:hto2gamma}.
\begin{align}
\Lambda(m_a^2;m_b,m_c)=&\frac{1}{m_a^2}\lambda^\frac{1}{2}(m_a^2,m_b^2,m_c^2)\mathrm{Log}\left(\frac{\lambda^\frac{1}{2}(m_a^2,m_b^2,m_c^2)-m_a^2+m_b^2+m_c^2}{2m_bm_c}\right)\,,\\[0.1cm]
C_0(m_Z^2,0,m_h^2,m_i^2,m_i^2,m_i^2)=&\frac{1}{2 \left(m_h^2-m_Z^2\right)}\left[\mathrm{Log}^2\left(\frac{\sqrt{-m_h^2 \left(4 m_i^2-m_h^2\right)}+2 m_i^2-m_h^2}{2 m_i^2}\right)\right.\notag\\
-&\left.\mathrm{Log}^2\left(\frac{\sqrt{-m_Z^2 \left(4 m_i^2-m_Z^2\right)}+2 m_i^2-m_Z^2}{2 m_i^2}\right)\right]\,,\\[0.1cm]
C_0(0,0,m_h^2,m_i^2,m_i^2,m_i^2)=&\frac{1}{2 m_h^2}\mathrm{Log}^2\left(\frac{\sqrt{-m_h^2 \left(4 m_i^2-m_h^2\right)}+2 m_i^2-m_h^2}{2 m_i^2}\right)\,,\\[0.1cm]
C_0(0,m_h^2,m_Z^2,m_i^2,m_i^2,m_j^2)=&\frac{1}{m_h^2-m_Z^2}\Bigg\{
-\mathrm{DiLog}\left[\frac{2(m_h^2+\Delta_{ij})}{m_h^2+\Delta_{ij}-\lambda^\frac{1}{2}(m_h^2,m_i^2,m_j^2)},m_h^2+\Delta_{ij}\right]\notag\\
-&\mathrm{DiLog}\left[\frac{2(m_h^2+\Delta_{ij})}{m_h^2+\Delta_{ij}+\lambda^\frac{1}{2}(m_h^2,m_i^2,m_j^2)},-(m_h^2+\Delta_{ij})\right]\notag\\
+&\mathrm{DiLog}\left[\frac{2\Delta_{ij}}{m_h^2+\Delta_{ij}-\lambda^\frac{1}{2}(m_h^2,m_i^2,m_j^2)},\Delta_{ij}\right]\notag\\
+&\mathrm{DiLog}\left[\frac{2\Delta_{ij}}{m_h^2+\Delta_{ij}+\lambda^\frac{1}{2}(m_h^2,m_i^2,m_j^2)},-\Delta_{ij}\right]-(m_h\to m_Z)\Bigg\}\,,
\end{align}
where $\Delta_{ij}=m_i^2-m_j^2$,
$\lambda(a,b,c)=a^2+b^2+c^2-2ab-2ac-2bc$ is the usual K{\"a}ll{\'e}n function, and $\mathrm{DiLog}[a,b]$ is a function defined in \textit{Package-X}~\cite{Patel:2015tea}.

\section{Kinematics}\label{app:kin}

Both $h(p)\to Z(p_1)\gamma(p_2)$ and $h(p)\to \gamma(p_1)\gamma(p_2)$ are processes of two-body decays, and its differential decay width is given by
\begin{align}\label{eq:dwidth}
d\Gamma=&\frac{1}{2m_h}|\mathcal{M}_X|^2d\Phi_2\,,
\end{align}
where $\mathcal{M}_X$ denotes the amplitude, and $d\Phi_2$ is the two-body decay phase space given by
\begin{align}
d\Phi_2=\frac{|\vec{p}_1|}{16\pi^2m_h}d\Omega\,.
\end{align}
Here, $|\vec{p}_1|=\lambda^{1/2}(m_h^2,m_1^2,m_2^2)/2m_h$, which is equal to $(m_h^2-m_Z^2)/2m_h$ for $h\to Z\gamma$ and $m_h/2$ for $h\to\gamma\gamma$, respectively; $d\Omega=d\phi_1 d\cos\theta_1$ is the solid angle of particle $1$ in the final state, which, after integration, yields $4\pi$. To calculate the decay width or branching ratio of the two processes, one has to sum over the spins of the $Z$ boson and photon in the final state:
\begin{align}\label{eq:sum}
\sum_{\lambda_1,\lambda_2}|\mathcal{M}_X|^2
=&|T_X|^2(p_2^\mu p_1^\nu-p_1\cdot p_2 g^{\mu\nu})(p_2^\alpha p_1^\beta-p_1\cdot p_2 g^{\alpha\beta})\notag\\
\times&\sum_{\lambda_1,\lambda_2}\left[\epsilon_\mu^\ast(p_1,\lambda_1)\epsilon_\alpha(p_1,\lambda_1)\right] \left[\epsilon_\nu^\ast(p_2,\lambda_2)\epsilon_\beta(p_2,\lambda_2)\right]\,.
\end{align}
For $X=Z\gamma$, one has
\begin{align}
\sum_{\lambda_1,\lambda_2}\left[\epsilon_\mu^\ast(p_1,\lambda_1)\epsilon_\alpha(p_1,\lambda_1)\right]\left[\epsilon_\nu^\ast(p_2,\lambda_2)\epsilon_\beta(p_2,\lambda_2)\right]=\left(-g_{\mu\alpha}+\frac{p_{1\mu}p_{1\alpha}}{m_Z^2}\right)(-g_{\nu\beta})\,.
\end{align}
Substituting this into Eqs.~\eqref{eq:sum} and~\eqref{eq:dwidth} and dividing by the total Higgs decay width $\Gamma_h$ yields the branching fraction for $h\to Z\gamma$,
\begin{align}\label{eq:BRhtoZgamma}
\mathcal{B}(h\to Z\gamma)=\frac{m_h^3}{32\pi\Gamma_h}\left(1-\frac{m_Z^2}{m_h^2}\right)^3|T_{Z\gamma}|^2\,.
\end{align}
Similarly, for $X=\gamma\gamma$, one has
\begin{align}
\sum_{\lambda_1,\lambda_2}\left[\epsilon_\mu^\ast(p_1,\lambda_1)\epsilon_\alpha(p_1,\lambda_1)\right]\left[\epsilon_\nu^\ast(p_2,\lambda_2)\epsilon_\beta(p_2,\lambda_2)\right]=\left(-g_{\mu\alpha}\right)(-g_{\nu\beta})\,,
\end{align}
and the branching fraction for $h\to\gamma\gamma$ reads
\begin{align}\label{eq:BRhto2gamma}
\mathcal{B}(h\to\gamma\gamma)=\frac{m_h^3}{64\pi\Gamma_h}|T_{\gamma\gamma}|^2 \,.
\end{align}
In deriving Eq.~\eqref{eq:BRhto2gamma}, since there are two identical photons in the final state, an additional $1/2$ factor is taken into account.

\bibliographystyle{apsrev4-1}
\bibliography{reference}

\end{document}